\documentclass[11pt,a4paper]{article}

\usepackage[T1]{fontenc}
\usepackage{newtxtext,newtxmath}
\usepackage[margin=1in]{geometry}

\usepackage{setspace}
\usepackage{titlesec}

\titleformat{\section}
{\normalfont\large\bfseries}
{\thesection.}{0.5em}{}

\titleformat{\subsection}
{\normalfont\normalsize\bfseries}
{\thesubsection.}{0.5em}{}

\titleformat{\subsubsection}
{\normalfont\normalsize\itshape}
{\thesubsubsection.}{0.5em}{}
\usepackage{amsmath}
\usepackage{graphicx}
\usepackage{booktabs}
\usepackage{tabularx}
\usepackage{array}
\usepackage{xcolor}
\usepackage{subcaption}
\usepackage{tikz}
\usetikzlibrary{positioning,arrows.meta,fit,backgrounds,calc}
\usepackage{pgfplots}
\pgfplotsset{compat=1.17}
\usepgfplotslibrary{groupplots}
\usepackage[numbers,sort&compress]{natbib}
\usepackage{microtype}
\usepackage[colorlinks=true,linkcolor=blue!50!black,citecolor=blue!50!black,urlcolor=blue!50!black]{hyperref}
\usepackage{doi}
\usepackage{cleveref}

\newcommand{\numHybCvsOF}{1.9\times10^{-7}}
\newcommand{\numHybVvsOF}{5.0\times10^{-10}}
\newcommand{\numMaxCdiff}{8.1\times10^{-13}}
\newcommand{\numMaxVdiff}{2.8\times10^{-12}}
\newcommand{\numMaxFieldDiff}{4.4\times10^{-16}}
\newcommand{\numTotalValues}{6.1}
\newcommand{\numTotalFields}{123}
\newcommand{\numOldCentreMax}{18}
\newcommand{\numOldCentreMed}{0.04}
\newcommand{\numOrdTwoDint}{2.00}
\newcommand{\numOrdTwoDbnd}{1.00}
\newcommand{\numOrdThreeDint}{1.97}
\newcommand{\numOrdThreeDintInf}{1.84}
\newcommand{\numOrdThreeDbnd}{1.01}
\newcommand{\numOrdTwoDall}{1.50}
\newcommand{\numOrdThreeDall}{1.48}
\newcommand{\numOrdWallCellTwo}{1.01}
\newcommand{\numOrdWallCellThree}{1.05}
\newcommand{\numOrdWallFixedThree}{0.99}
\newcommand{\numWallFixedTwoMax}{5.0\times10^{-11}}
\newcommand{\numCellExportMax}{4.2\times10^{-15}}
\newcommand{\numLoadPerFace}{1.3}
\newcommand{\numBigCells}{262{,}144}
\newcommand{\numBigLoad}{1.3}
\providecommand{\numRestyleMax}{359}
\providecommand{\numNewFieldMax}{305}
\providecommand{\numProdLoadMin}{0.268}
\providecommand{\numProdLoadMax}{0.433}

\newcommand{\HP}{HAVAPlot}
\newcommand{\vect}[1]{\mathbf{#1}}
\newcolumntype{H}{>{\setbox0=\hbox\bgroup}c<{\egroup}@{}}

\title{\textbf{\HP: A Web-Based Scientific Visualization and Post-Processing Platform}}

\author{
\textbf{Ozgur Tumuklu}\thanks{Corresponding author: \href{mailto:tumuko@rpi.edu}{tumuko@rpi.edu}}\\[1.5ex]
{\small HAVA Laboratory, Department of Mechanical, Aerospace, and Nuclear Engineering}\\
{\small Rensselaer Polytechnic Institute, Troy, NY 12180, USA}
}

\date{}

\begin{document}
\maketitle

\begin{abstract}
This paper presents HAVAPlot, a browser-based framework for scientific visualization and post-processing of numerical simulation data. Operating entirely within the user's browser, HAVAPlot eliminates software installation and external data transfers while providing interactive analysis and publication-quality graphical output. Although primarily developed for computational fluid dynamics, the framework supports mesh-based and tabular datasets from other disciplines. HAVAPlot directly reads native OpenFOAM cases, including polyhedral meshes, ASCII and binary fields, boundary data, and Lagrangian particles, as well as legacy and XML VTK files. Its capabilities include two- and three-dimensional visualization, exact polyhedral slicing, boundary data extraction, mathematical field operations, streamlines, isosurfaces, particle visualization, customizable plotting, and vector PDF/EPS export. The underlying numerical algorithms are independently verified. Cell centers and volumes agree with an independent implementation of OpenFOAM's geometric algorithm, while 6.1 million field values across four production cases are reproduced to machine precision. Analytical tests on distorted hexahedral meshes demonstrate second-order interior and first-order boundary convergence for cell-to-point interpolation. Meshes containing 250,000 cells load in at most 1.3 seconds. Applications spanning supersonic and hypersonic flows, hybrid particle-continuum simulations, and gas-bearing configurations demonstrate the framework's versatility. HAVAPlot provides an accessible, installation-free environment for scientific data analysis and visualization.
\end{abstract}


\section{Introduction}
\label{sec:intro}

Post-processing is an essential component of computational research, enabling the interpretation, quantitative analysis, and visualization of numerical simulation results across disciplines such as computational fluid dynamics, structural mechanics, and heat transfer. A typical session consists of opening a solution, inspecting a few fields, cutting planes, extracting values along walls or lines, comparing cases or time steps, and, finally, preparing figures for a report or a paper. General-purpose visualization platforms such as ParaView~\cite{ahrens2005,ayachit2015} and VisIt~\cite{childs2012}, which leverage the Visualization Toolkit (VTK)~\cite{schroeder2006}, offer extensive capabilities that extend well beyond these requirements. Similarly, scripting libraries such as Matplotlib~\cite{hunter2007} and PyVista~\cite{sullivan2019} provide flexible and highly customizable visualization capabilities for users with programming expertise. Despite their extensive capabilities, these tools generally require local installation, dependency management, or programming expertise, which can limit their accessibility when software installation is restricted or rapid visualization is required.

Web-based visualization has advanced considerably, but web tools usually require the data to be converted to a web format or uploaded to a server, and they are designed for display rather than for quantitative post-processing.
\HP\ addresses these limitations by providing an installation-free framework that integrates native simulation-data reading, quantitative post-processing, and publication-quality scientific visualization within a standard web browser. It is written in JavaScript, runs entirely on the user's computer, and is designed for current versions of the major web browsers. The tests reported in this paper were run in Google Chrome. Saving files directly into a folder chosen by the user requires the File System Access API, which is available in Chromium-based browsers; other browsers save the files as downloads. The user selects an OpenFOAM~\cite{weller1998} case, VTK files~\cite{vtkformats} or tables, and all subsequent steps, namely parsing, geometry, interpolation, rendering and export, are carried out on the user's own computer, so that large or confidential simulation files are never transferred to an external server. Because VTK is a common exchange format of many finite-volume and finite-element codes, the same workflow applies to results from structural and thermal solvers as well as to those of flow solvers. The tool is part of the online tools of the HAVA Laboratory at \url{https://www.havalab.org/handy-tools/} and is free to use.

The contributions of this paper are as follows:
\begin{enumerate}
\item The design of a complete post-processing pipeline that runs in the browser, from reading polyhedral meshes, boundary data and Lagrangian clouds to vector-graphics output (\cref{sec:design,sec:reading}).
\item The algorithms that make the results quantitatively correct: solver-consistent cell geometry, cell-to-point interpolation, exact slicing of polyhedral cells with second-order point values, data on boundary patches, and color scales with linear, exponential or user-defined levels (\cref{sec:algorithms}).
\item A verification of these algorithms against an independent reader, which implements OpenFOAM's geometric algorithm, on four production cases, and against analytical test fields on distorted meshes, from which the observed orders of accuracy are obtained, together with performance measurements (\cref{sec:verification}).
\item Applications to compressible, hypersonic, multiphase and three-dimensional flows (\cref{sec:applications}).
\end{enumerate}
The verification scripts, the generator of the synthetic test cases and the raw results are provided as ancillary files.

\section{Design and architecture}
\label{sec:design}

\subsection{Goals}
Four goals guided the design. (i)~\emph{No installation and no upload.} The tool is a static web page, and files are read through the browser's File API without being transmitted. In browsers that support the File System Access API~\cite{whatwg_fs,wicg_fsaccess}, figures, layouts, and data can also be saved directly to a folder of the user's choice, such as the folder of the OpenFOAM case~\cite{weller1998} being analyzed. (ii)~\emph{Native input.} Solver output is read in its native format, for example, an OpenFOAM case exactly as written by the solver, so that no conversion step (such as OpenFOAM's \texttt{foamToVTK} utility) is required. (iii)~\emph{Quantitative correctness.} Every value that is displayed or exported must be either the solver's value or a well-defined interpolation of it. (iv)~\emph{Publication-quality output.} A single drawing routine produces the screen image, PNG at a chosen resolution, and vector EPS and PDF files with \LaTeX-style labels.

\subsection{Architecture}
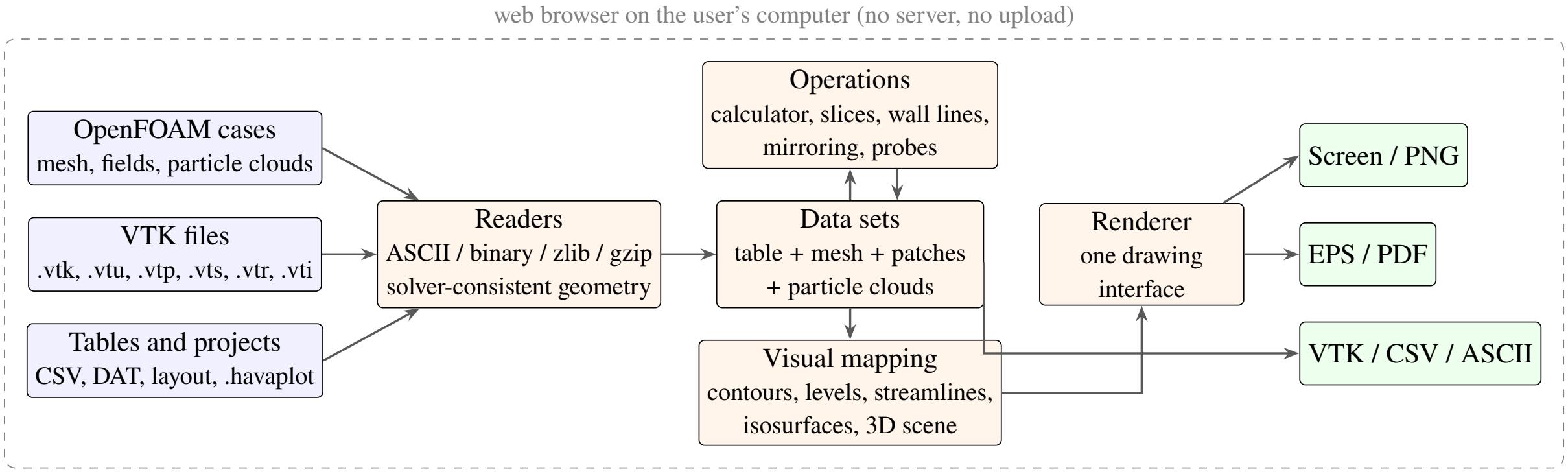
\begin{figure}[!htbp]
\centering
\resizebox{\textwidth}{!}{%
\begin{tikzpicture}[
  font=\small, node distance=4mm and 7mm,
  box/.style={draw, rounded corners=2pt, align=center, minimum height=8mm, inner sep=3pt, fill=white},
  ibox/.style={box, fill=blue!6}, pbox/.style={box, fill=orange!8}, obox/.style={box, fill=green!7},
  arr/.style={-{Stealth[length=2mm]}, thick, gray!70!black}]
\node[ibox] (of) {OpenFOAM cases\\\footnotesize mesh, fields, particle clouds};
\node[ibox, below=of of] (vtk) {VTK files\\\footnotesize .vtk, .vtu, .vtp, .vts, .vtr, .vti};
\node[ibox, below=of vtk] (tab) {Tables and projects\\\footnotesize CSV, DAT, layout, .havaplot};
\node[pbox, right=of vtk, minimum width=3.3cm] (rd) {Readers\\\footnotesize ASCII / binary / zlib / gzip\\\footnotesize solver-consistent geometry};
\node[pbox, right=of rd, minimum width=3.4cm] (ds) {Data sets\\\footnotesize table $+$ mesh $+$ patches\\\footnotesize $+$ particle clouds};
\node[pbox, above=of ds, minimum width=3.4cm] (op) {Operations\\\footnotesize calculator, slices, wall lines,\\\footnotesize mirroring, probes};
\node[pbox, below=of ds, minimum width=3.4cm] (vis) {Visual mapping\\\footnotesize contours, levels, streamlines,\\\footnotesize isosurfaces, 3D scene};
\node[pbox, right=of ds, minimum width=2.6cm] (ren) {Renderer\\\footnotesize one drawing\\\footnotesize interface};
\node[obox, above right=2mm and 7mm of ren] (scr) {Screen / PNG};
\node[obox, right=7mm of ren] (vec) {EPS / PDF};
\node[obox, below right=2mm and 7mm of ren] (dat) {VTK / CSV / ASCII};
\draw[arr] (of.east) -- (rd); \draw[arr] (vtk) -- (rd); \draw[arr] (tab.east) -- (rd);
\draw[arr] (rd) -- (ds);
\draw[arr] (ds.north) -- (op.south); \draw[arr] ([xshift=6mm]op.south) -- ([xshift=6mm]ds.north);
\draw[arr] (ds) -- (vis);
\draw[arr] (vis.east) -| (ren.south); \draw[arr] (ren) -- (scr.west); \draw[arr] (ren) -- (vec); \draw[arr] (ds.east) ++(0,-1mm) |- (dat.west);
\begin{scope}[on background layer]
\node[draw, dashed, rounded corners=4pt, gray, fit=(of)(tab)(op)(vis)(ren)(scr)(dat), inner sep=8pt, label={[gray,font=\footnotesize]above:web browser on the user's computer (no server, no upload)}] {};
\end{scope}
\end{tikzpicture}}
\caption{Structure of \HP. Every input becomes a data set (a table with an optional mesh, its boundary patches and particle clouds); slices, wall lines and calculated variables are data sets as well. A single drawing interface serves the screen, raster and vector outputs.}
\label{fig:arch}
\end{figure}

The overall structure that follows from these goals is shown in \cref{fig:arch}. Its central element is a common \emph{data set}, into which the readers convert every input. Each data set is organized as a table, with rows representing individual data points and columns representing their associated variables. The first columns contain the spatial coordinates, followed by additional fields such as scalar or vector quantities. Depending on the input, a data set may also include mesh connectivity information (e.g., 2D polygonal cells, polyhedral meshes, or VTK cells), boundary patches with associated field values, or particle clouds. Vector fields are stored as one column per component plus a column for the magnitude. Since all subsequent operations act on data sets and also return data sets, a plane slice or a line along a wall can itself be plotted, combined with other data, used in the calculator or exported, exactly like the original input.

The final stage of the pipeline, the renderer, is deliberately kept small. It is an interface consisting of paths, filled shapes, text and raster images, with three implementations: the HTML canvas for the screen and PNG output, and EPS and PDF writers. Because all three share the same drawing calls, labels containing \LaTeX-style mathematics are typeset identically on screen and in the vector files. Smooth color shading is embedded as a raster image in the vector files, whereas filled bands, lines, text and markers remain vector elements.

\begin{figure}[!htbp]
\centering
\includegraphics[width=\textwidth]{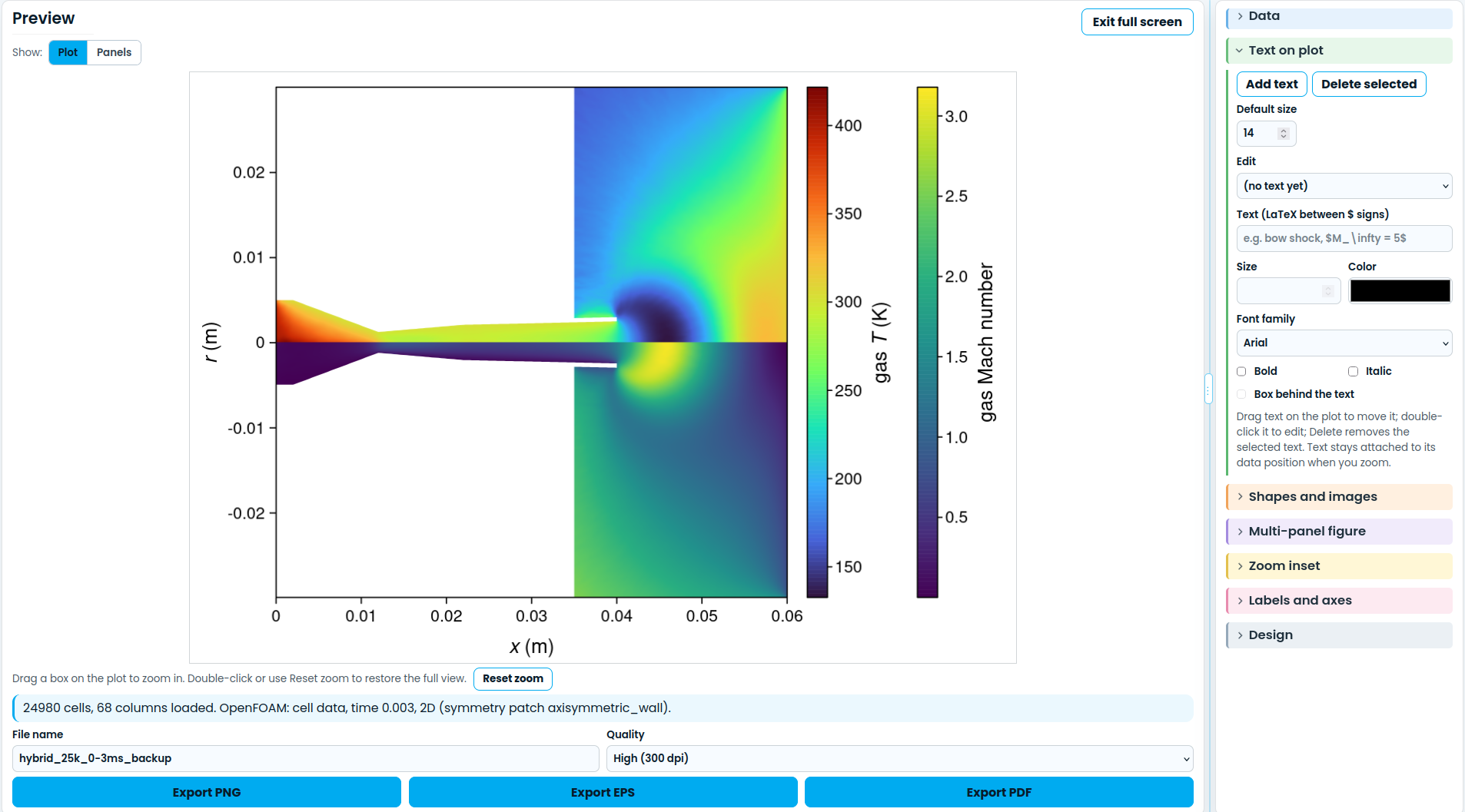}
\caption{The \HP\ interface with the axisymmetric cold-spray case of \cref{sec:hybrid} open: preview (left; gas temperature and gas Mach number) and the color-coded settings panels (right; the divider between them can be dragged). The file row shows the case, the number of cells and the time.}
\label{fig:ui}
\end{figure}

\HP is implemented in JavaScript using three principal modules: an OpenFOAM reader (\texttt{foam-reader.js}), a VTK reader (\texttt{vtk-reader.js}), and the visualization and post-processing module (\texttt{plot-tools.js}). It has no external dependencies; decompression, for instance, relies on the browser's built-in \texttt{DecompressionStream}. The user interface is shown in \cref{fig:ui}. The settings are arranged in collapsible, color-coded panels (data, text on plot, shapes and images, multi-panel figures, zoomed insets, colorbar, design), so that options belonging together are visually grouped and a new user can find them without consulting a manual. All settings of a plot can be stored as a small layout file and applied to another case, or saved together with the data as a single compressed project file, which reopens the complete plot without the original files.

\section{Reading data}
\label{sec:reading}

\subsection{OpenFOAM cases}
The user selects the case folder or drops it onto the page. \HP\ then lists the available output times and reads the polyhedral mesh, that is, the point coordinates, the faces as lists of points, the owner--neighbour connectivity of the faces and the definition of the boundary patches, together with every volume field of the chosen time, in ASCII or binary format and optionally gzip-compressed. Uniform fields are expanded, and boundary values that the solver writes explicitly, either uniform or as non-uniform lists, are retained for every patch, including patches selected by regular expressions. Further output times are read on request and reuse the mesh. Several cases can be open simultaneously, for example to compare two runs along the same wall. Cases that exist only in decomposed form are recognised, and the user is asked to reconstruct them first.

\paragraph{Cell geometry.} Accurate post-processing requires a consistent geometric representation of the computational mesh, particularly when associating cell-centred values with their spatial locations. HAVAPlot therefore computes face centres, face area vectors, cell centres, and cell volumes using the same geometric decomposition as OpenFOAM. Consider a face $f$ with $m$ vertices at positions $\vect{p}_1,\dots,\vect{p}_m$, numbered around the face (with $\vect{p}_{m+1}=\vect{p}_1$), and let $\bar{\vect{p}}=\frac1m\sum_i\vect{p}_i$ be their mean. The face is divided into the $m$ triangles $(\vect{p}_i,\vect{p}_{i+1},\bar{\vect{p}})$, $i=1,\dots,m$, which have the normals $\vect{n}_i=(\vect{p}_{i+1}-\vect{p}_i)\times(\bar{\vect{p}}-\vect{p}_i)$ (twice their area vectors) and the centroids $\vect{c}_i=(\vect{p}_i+\vect{p}_{i+1}+\bar{\vect{p}})/3$. With $\vect{N}=\sum_i\vect{n}_i$ and the weights $a_i=\vect{n}_i\cdot\vect{N}/|\vect{N}|$, the face area vector $\vect{S}_f$ (normal to the face, with magnitude equal to its area) and the face centre $\vect{x}_f$ are
\begin{equation}
\vect{S}_f=\tfrac12\vect{N},\qquad \vect{x}_f=\frac{\sum_i a_i\vect{c}_i}{\sum_i a_i}.
\end{equation}
Each cell $c$ is then split into pyramids, one per face $f\in c$ (the faces bounding the cell), which join the face to an estimated cell centre $\tilde{\vect{x}}_c$, the mean of the cell's face centres. The volume of such a pyramid is $V_{cf}=\pm\frac13\vect{S}_f\cdot(\vect{x}_f-\tilde{\vect{x}}_c)$, where the sign is positive if $\vect{S}_f$ points out of cell $c$ and negative otherwise. The cell volume $V_c$ is the sum of the pyramid volumes, and the cell centre $\vect{x}_c$ is the volume-weighted mean of the pyramid centroids, which lie at three quarters of the distance from the apex to the base:
\begin{equation}
V_c=\sum_{f\in c}V_{cf},\qquad \vect{x}_c=\frac{1}{V_c}\sum_{f\in c}V_{cf}\left(\tfrac34\vect{x}_f+\tfrac14\tilde{\vect{x}}_c\right).
\label{eq:centroid}
\end{equation}
The simpler choice of the estimate $\tilde{\vect{x}}_c$ itself as the cell centre is not sufficient: in the degenerate wedge cells adjacent to the axis of axisymmetric meshes, it deviates from $\vect{x}_c$ by up to \numOldCentreMax\% of the cell size (median \numOldCentreMed\%; \cref{sec:reader}), which is why the exact form \eqref{eq:centroid} is used.

\paragraph{Two-dimensional and axisymmetric cases.} OpenFOAM represents 2D and axisymmetric problems by meshes that are one cell thick. \HP\ recognises such meshes from patches of type \texttt{empty} or \texttt{wedge}; failing that, it uses a patch with two faces on every cell (axisymmetric meshes whose front and back form one \texttt{symmetry} patch) or a pair of patches with one face on every cell. The thin direction is then taken as the axis along which these faces lie farthest from the cell centres, and the polygon of each cell is its side face lowest along that axis, so that all polygons are taken from the same side. A case recognised in this way is plotted on its own cell polygons (\cref{sec:interp}), although the user may still open it as 3D.

\paragraph{Lagrangian particles.} Particle clouds stored with a case are detected automatically. Particle positions are taken from Cartesian coordinates when these are written; a cloud stored only in barycentric (tetrahedral) coordinates is instead placed at the centres of the particles' cells and marked as approximate. Every per-particle property, whether scalar, vector (stored as components and magnitude) or integer label, becomes a column, whereas properties stored as a list per particle are skipped. Because clouds can be very large, the number of particles in each cloud is shown before it is read, and a cloud can be thinned to a chosen number of particles while it is read. The particles are then drawn over the 2D plot or in the 3D scene, colored and sized by any property, filtered by a property range or by their distance from a slice plane, and inspected individually with the probe.

\subsection{VTK files and tables}
The VTK reader provides the route by which data from other solvers, including finite-element codes for structural and thermal analysis, enter \HP. It handles legacy files (ASCII and binary; structured points, structured and rectilinear grids, polydata and unstructured grids) as well as the XML formats \texttt{.vti}, \texttt{.vtr}, \texttt{.vts}, \texttt{.vtu} and \texttt{.vtp} with ASCII, inline binary or appended raw/base64 data, uncompressed or zlib-compressed, and with 32- or 64-bit headers. Point data are used when present; files with cell data only are tabulated at the cell centres and retain their 2D cell polygons. Finally, comma- or space-separated tables and zoned ASCII data files are read with automatic detection of separators, headers and zones.

\section{Algorithms}
\label{sec:algorithms}

\subsection{Cell data on two-dimensional (2D) meshes}
\label{sec:interp}
A finite-volume solver stores one value per cell, which represents the average over the cell and is associated with its centre. In order to draw smooth contours, these cell values are first transferred to the mesh points by the arithmetic mean of the cells surrounding each point,
\begin{equation}
\phi_p=\frac{1}{|\mathcal{C}(p)|}\sum_{c\in\mathcal{C}(p)}\phi_c,
\label{eq:c2p}
\end{equation}
where $\phi$ denotes any scalar field of the data set (for example the pressure, the temperature or one component of the velocity), $\phi_c$ its value in cell $c$, $\phi_p$ the resulting value at mesh point $p$, $\mathcal{C}(p)$ the set of cells that have $p$ as a vertex, and $|\mathcal{C}(p)|$ the number of these cells. This is also the operation performed by ParaView's \emph{Cell Data to Point Data} filter. For the smoothly distorted meshes considered in the present verification study, this averaging procedure exhibits second-order convergence at interior points and first-order convergence at boundary points, where the average is one-sided (\cref{sec:verification}). Second-order accuracy in the interior is not guaranteed on arbitrary meshes, however: it requires the mesh point to lie close to the centroid of the surrounding cell centres, as on smooth structured or nearly uniform meshes. On strongly stretched or irregular unstructured meshes, the arithmetic mean may be only first-order accurate. Each cell polygon is subsequently split into a fan of triangles, and the point values are interpolated linearly onto a regular sampling grid whose resolution follows the number of cells and is refined when the user zooms in. Sample points that lie outside every polygon remain blank, so that holes such as solid bodies in CFD stay empty. To keep curved boundaries smooth on coarse meshes, the outline of the mesh is, in addition, extracted (edges used by only one polygon, chained into loops) and the fill is clipped to it.

Data that are not of this form are handled analogously: scattered data without a mesh are gridded by inverse-distance weighting, structured grids (including curvilinear ones) are interpolated linearly on their own cells, and complete coordinate lattices are used directly.

\subsection{Filled contours and color scales}
\label{sec:levels}
Contour bands are drawn by splitting every sampling cell into two triangles, within which the field is linear, and by filling, for each level $L_k$ in increasing order, the region where $\phi\ge L_k$. Drawing one filled shape per band in this way avoids hairline seams between adjacent pieces. Contour lines are obtained by marching squares on the same grid. In the smooth mode, every output pixel is interpolated within the same triangles and colored through a 1\,024-entry look-up table, whereas a ``soft bands'' mode retains the band colors but blends them over part of each band.

The color scale spans the range $[\phi_{\min},\phi_{\max}]$, which is the range of the data by default and may be set by the user, and is divided into $n$ bands by the levels $L_0=\phi_{\min}<L_1<\dots<L_n=\phi_{\max}$. A value $\phi$ is mapped to the position $t(\phi)\in[0,1]$ along the colormap, from its first color ($t=0$) to its last ($t=1$), such that the levels are equally spaced in $t$, i.e.\ $t(L_k)=k/n$:
\begin{equation}
t(\phi)=\begin{cases}
\dfrac{\phi-\phi_{\min}}{\phi_{\max}-\phi_{\min}} & \text{linear},\\[2ex]
\dfrac{\log\phi-\log\phi_{\min}}{\log\phi_{\max}-\log\phi_{\min}} & \text{exponential},\\[2ex]
\dfrac{1}{n}\left(k+\dfrac{\phi-L_k}{L_{k+1}-L_k}\right),\ L_k\le\phi<L_{k+1} & \text{manual levels } L_0<\dots<L_n.
\end{cases}
\label{eq:levels}
\end{equation}
Exponential levels thus form a geometric sequence, $L_k=\phi_{\min}(\phi_{\max}/\phi_{\min})^{k/n}$, and values at or below zero take the lowest color. With manual levels, each band receives an equal share of the colormap irrespective of its width in $\phi$. The same function $t$ positions the numbers on the colorbar, which may be chosen automatically (round numbers, or decades on the exponential scale), placed at every level, or given as a list. Since bands, lines, colorbar and 3D colors all use \eqref{eq:levels}, they are consistent by construction. Perceptually uniform colormaps (viridis, cividis~\cite{nunez2018}, Turbo~\cite{mikhailov2019}, a diverging map~\cite{moreland2009} and others) are provided; the well-known shortcomings of the traditional rainbow map~\cite{borland2007} motivated the choice of Turbo as the default.

\subsection{Plane slices of polyhedral meshes}
\label{sec:slice}
For a plane through the point $\vect{o}$ with unit normal $\vect{n}$, the signed distance $d_p=(\vect{x}_p-\vect{o})\cdot\vect{n}$ from the plane is computed once for every mesh point $p$ at position $\vect{x}_p$. Every face edge joining two mesh points $p$ and $q$ with $d_p$ and $d_q$ of opposite sign is crossed by the plane, at the cut point
\begin{equation}
\vect{x}=\vect{x}_p+s\,(\vect{x}_q-\vect{x}_p),\qquad s=\frac{d_p}{d_p-d_q},
\end{equation}
where $s\in[0,1]$ is the fraction of the edge length from $p$ to the cut point. The cut point is shared by all cells around the edge. For convex polyhedral cells, the cut points, ordered by angle around their mean in the plane, define the intersection polygon. This construction locates the intersection points directly on the original mesh edges, without volumetric resampling. The slice is a 2D data set of polygons that carries the values of the cut cells and is drawn with the methods of \cref{sec:interp}. The value at a polygon corner $\vect{x}$ is obtained from the values $\phi_p$ and $\phi_q$ at the two ends of the cut edge, given by \eqref{eq:c2p}, by linear interpolation along the edge,
\begin{equation}
\phi(\vect{x})=(1-s)\,\phi_p+s\,\phi_q ,
\label{eq:cutvalue}
\end{equation}
which retains the accuracy of the point values \eqref{eq:c2p}: on the test meshes of \cref{sec:mms}, the slice values are second-order accurate in the interior and first-order accurate at the boundary. A plane that lies exactly on the outer boundary of the mesh is moved inwards by $10^{-6}$ of the domain size so that it intersects the boundary cells. Planes may be normal to a coordinate axis or arbitrarily oriented, and several planes can be retained and shown together in the 3D view, each with its own variable, colormap and levels.

\subsection{Data on boundary patches}
\label{sec:surface}
Wall quantities such as heat flux, pressure or shear stress are frequently written only on boundary patches, so that they cannot be recovered from the cell values. For a chosen patch, \HP\ therefore either lists all of its faces (face centre and values) or extracts a line. In the latter case, the patch faces crossed by a plane give segments, which are chained into polylines through their shared cut points (points at the same position, such as on an axis, are joined); each segment then becomes a row at its midpoint with the arc length $s$. The cutting plane is the mid-plane of the case for 2D problems, the current slice, or a plane $x$, $y$ or $z=\text{const}$. The values are the field's own boundary values on that patch wherever the solver provides them (for example a wall heat flux or a prescribed wall temperature) and otherwise the values of the wall-adjacent cells; the tool reports which of the two applies to each field. In 3D, the patches themselves can also be shown in the 3D view, colored by their face values.

\subsection{Three-dimensional (3D) view}
The 3D scene, comprising slice planes, domain walls or selected patches, isosurfaces, streamlines, particles and axes, is drawn with an affine projection and the painter's algorithm~\cite{newell1972}. All primitives are sorted by depth and drawn from back to front with the 2D renderer; as a result, the 3D view can be exported as vector graphics in the same way as a 2D plot. Faces of walls that point away from the viewer are culled. Isosurfaces are extracted by marching tetrahedra~\cite{doi1991} on a sampling lattice, with six tetrahedra per cell split identically in every cell so that the surfaces close; this choice avoids the topological ambiguities of marching cubes~\cite{lorensen1987}. Contour lines on slice polygons are extracted, analogously, by marching triangles on the fan triangulation of each polygon. The view is rotated, panned and zoomed with the mouse, and axis titles may contain \LaTeX\ and are drawn in perspective.

\subsection{Streamlines, calculator and probes}
Streamlines in a plane are integrated with a second-order (midpoint) Runge--Kutta scheme on the normalised velocity field. They are started either at points selected by the user or automatically. For the automatic seeding, a coarse grid is laid over the plot, whose resolution is set by the chosen line density; a streamline is started from every other cell of this grid unless an earlier streamline has already passed through that cell. This is a simplified form of the method of Jobard and Lefer~\cite{jobard1997}: it avoids duplicated and crowded lines, but, unlike the original method, it does not stop a line where it approaches another one, so that lines may still come close together where the flow converges. In 3D, streamlines can also be traced through the volume. The variable calculator defines new columns by expressions of existing ones, such as the Mach number in \cref{sec:jet}. It uses its own parser rather than JavaScript evaluation, supports the usual mathematical functions as well as whole-column statistics such as $\max(x)$, and recreates the variables automatically when another time step is read. Finally, the probe reports the values under the cursor: the series of a line plot where the cursor crosses them, the field (or all fields) of a contour plot, the values on slice planes and faces in 3D, and the properties of a particle.

\subsection{Writing data}
The processed data, whether a loaded data set, a slice or a particle cloud, can be written together with any calculated variables as legacy VTK files (polydata or structured grid), as zoned ASCII finite-element files in which polygons are stored as quadrilateral elements, or as comma- or space-separated tables. The mesh formats can store the values either at the cells, which is exact, or at the mesh points according to \cref{eq:c2p,eq:cutvalue}. Cropping to the zoomed region and the number of significant digits (up to 17, i.e.\ full double precision) are available as options. Written files reopen in \HP\ and in other programs on the same polygons; with 17 significant digits, the cell values are written and read back exactly (\cref{sec:roundtrip}).

\section{Verification and performance}
\label{sec:verification}

All tests were run in Google Chrome~156 (headless) on Linux, on an AMD Ryzen~7 5700G with 64~GB of memory. The tests are driven by scripts that operate the actual user interface through the Chrome DevTools Protocol: files are chosen in the same file inputs that a user would use, controls are set, and the exported files are captured. All results are compared with independent references rather than with output of \HP\ itself. The interpolation tests (\cref{sec:mms}) follow the idea of the method of manufactured solutions, which is widely used to verify solvers~\cite{roache2002}: an analytical field, whose exact value is known at every point, is prescribed on a sequence of refined meshes, and the error of the computed result is measured against it. The rate at which this error decreases with the cell size gives the observed order of accuracy, which can then be compared with the theoretical one.

\subsection{Reader and geometry}
\label{sec:reader}
\Cref{tab:reader} lists the four production cases used for this purpose. They were computed with OpenFOAM-based solvers and were chosen to cover different mesh types, physical models and output formats. The first is an axisymmetric cold-spray nozzle computed with an in-house hybrid particle solver of the HAVA Laboratory (\texttt{uniGasHAVAMPPICFoam}), in which the gas is described by the direct simulation Monte Carlo (DSMC) method~\cite{bird1994} and the solid particles by the multiphase particle-in-cell (MP-PIC) method~\cite{andrews1996}; its output contains a large number of fields, including statistical moments of the gas and two particle clouds. The second is an axisymmetric underexpanded supersonic jet impinging on a flat plate, computed with OpenFOAM's density-based solver \texttt{rhoCentralFoam}~\cite{greenshields2010}. The third is the hypersonic flow of air over a circular cylinder at conditions of the HEG shock tunnel~\cite{knight2012}, computed with the two-temperature solver \texttt{hy2Foam}~\cite{casseau2016} with finite-rate chemistry, which requires a separate field for every species and both temperatures. The fourth is the three-dimensional gas film of a self-acting slider bearing, computed with OpenFOAM's steady compressible solver \texttt{rhoSimpleFoam}; it is the only case with a fully three-dimensional mesh. To provide an independent reference, a Python reader was written for this test. It parses every field file and computes cell centers and volumes with OpenFOAM's algorithm. For the cold-spray case, for which the solver had also written its own cell centers and volumes, the Python geometry reproduces these to $\numHybCvsOF$ of the cell size and $\numHybVvsOF$ relative, which corresponds to the precision of the written files.

The geometry computed in the browser agrees with this reference to within $\numMaxCdiff\,h$ for the centers ($h=V_c^{1/3}$) and $\numMaxVdiff$ relative for the volumes, that is, to rounding error. Likewise, all \numTotalFields\ field columns, \numTotalValues~million values in total, agree with the independent parse. The values read from the files are bitwise identical (checked for the first three cases); the only differences, at most $\numMaxFieldDiff$ relative, that is, $2\varepsilon$ in double precision, occur in the vector magnitudes, which both readers compute from the components. The last column of the table, finally, quantifies the error that the simpler center estimate $\tilde{\vect{x}}_c$ would introduce; it is largest in the two axisymmetric cases, whose wedge cells degenerate on the axis.

\begin{table}[!htbp]
\centering
\caption{Production cases read by \HP\ and compared with an independent reader. Columns: field columns compared (a vector field contributes its three components and its magnitude); $\Delta x_c$: largest distance between \HP's cell centres and the reference, divided by $h=V_c^{1/3}$; $\Delta V$, $\Delta\phi$: largest relative differences of cell volumes and of all field values; last column: the error of the mean of the face centres used as the centre. Load: time to read the mesh and all fields of one time step.}
\label{tab:reader}
\small
\setlength{\tabcolsep}{3.5pt}
\resizebox{\textwidth}{!}{%
\begin{tabular}{@{}lH r r r r c c c c@{}}
\toprule
Case & & Cells & Columns & Values & Load (ms) & $\Delta x_c/h$ & $\Delta V$ & $\Delta\phi$ & $|\tilde{\vect{x}}_c-\vect{x}_c|/h$ \\
\midrule
\csname @@input\endcsname data/reader_table.tex
\bottomrule
\end{tabular}}
\end{table}

\subsection{Manufactured fields on distorted meshes}
\label{sec:mms}
To quantify the interpolation accuracy, OpenFOAM cases were generated on the unit square (one cell thick, with \texttt{empty} front and back) and on the unit cube, with $N^2$ and $N^3$ hexahedral cells, respectively, where $N$ is the number of cells per direction and $h=1/N$ the nominal cell size. The mesh points of a uniform mesh, at $\boldsymbol{\xi}=(\xi,\eta,\zeta)\in[0,1]^3$, were moved to the positions
\begin{equation}
\vect{x}=\boldsymbol{\xi}+a\,\sin(2\pi\xi)\sin(2\pi\eta)\,[\sin(\pi\zeta)]\,(1,1,[1]),\qquad a=0.03,
\end{equation}
where $\vect{x}=(x,y,z)$, $a$ is the amplitude of the distortion, and the bracketed factors apply in 3D only. The distortion vanishes on the boundary, so that the domain itself is unchanged. The cells carry the exact values of the test fields
\begin{align}
\phi_{2D}&=\sin(2\pi x)\cos(\pi y)+\tfrac12xy, \\
\phi_{3D}&=\sin(\pi x)\cos(\pi y)\,e^{z}+0.3\,xyz
\end{align}
at their centroids \eqref{eq:centroid}. These tests therefore assess interpolation and slicing for centroid-sampled fields, rather than for exact cell-volume averages. Each case contains the test field twice, with the same cell values: as $\phi$, whose exact values on the bottom wall ($y=0$) are also written as a non-uniform \texttt{fixedValue} boundary condition, and as $\phi^\ast$, which has only \texttt{zeroGradient} conditions. In this way, the two possible sources of wall data in \cref{sec:surface}, the boundary values written by the solver and the values of the wall-adjacent cells, can be tested separately. The cases were opened through the user interface, and in 3D an oblique slice through $(\frac12,\frac12,\frac12)$ with normal $(1,1,1)/\sqrt3$ was taken. The mesh-point values were then exported as VTK, the cell values as CSV, and the bottom-wall line (2D mid-plane; plane $z=\frac12$ in 3D) as CSV, and the exported values were compared with the exact field at the exported positions. For a set of $M$ points (for example the interior mesh points) with exported values $\phi_j$ at the positions $\vect{x}_j$, the error at each point is $e_j=|\phi_j-\phi_{\mathrm{exact}}(\vect{x}_j)|$, and two measures of the error are used: the root-mean-square (RMS) error and the maximum error,
\begin{equation}
E_{\mathrm{RMS}}=\Bigl(\frac1M\sum_{j=1}^{M}e_j^2\Bigr)^{1/2},\qquad E_{\max}=\max_{1\le j\le M} e_j .
\end{equation}
The RMS error measures the typical error over the set, whereas the maximum error (maximum norm) is determined by the worst point. If the error behaves as $E\propto h^{q}$, then $q$ is the order of accuracy. The observed order between two successive meshes with $N_1<N_2$ cells per direction is $q=\ln(E_1/E_2)/\ln(N_2/N_1)$, and the values quoted below are the means of $q$ over the three pairs of finest meshes.

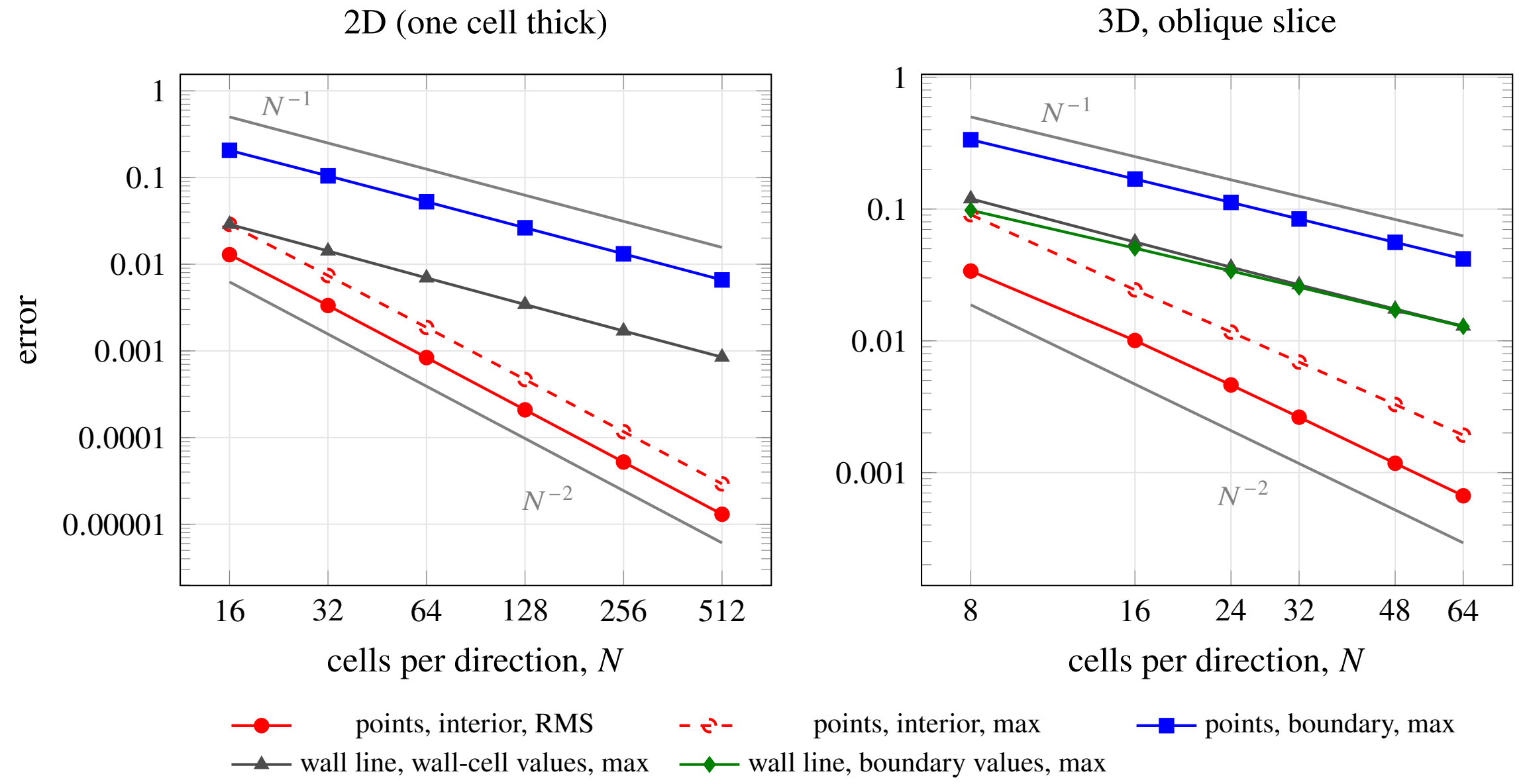
\begin{figure}[!htbp]
\centering
\begin{tikzpicture}
\begin{groupplot}[group style={group size=2 by 1, horizontal sep=1.5cm},
  width=0.47\textwidth, height=0.42\textwidth, xmode=log, ymode=log, log basis x=2,
  xtick=data, log ticks with fixed point, xlabel={cells per direction, $N$}, grid=major, grid style={gray!20},
  legend columns=3, legend style={font=\scriptsize, draw=none, /tikz/every even column/.append style={column sep=6pt}},
  every axis plot/.append style={thick, mark size=1.8pt}, tick label style={font=\footnotesize}, label style={font=\small}, title style={font=\small}]
\nextgroupplot[title={2D (one cell thick)}, ylabel={error}, legend to name=convleg]
\addplot[red, mark=*] table[x=N, y=nodeint_l2] {data/conv2d.dat}; \addlegendentry{points, interior, RMS}
\addplot[red, dashed, mark=o] table[x=N, y=nodeint_linf] {data/conv2d.dat}; \addlegendentry{points, interior, max}
\addplot[blue, mark=square*] table[x=N, y=node_bnd_linf] {data/conv2d.dat}; \addlegendentry{points, boundary, max}
\addplot[black!70, mark=triangle*] table[x=N, y=wall_cell] {data/conv2d.dat}; \addlegendentry{wall line, wall-cell values, max}
\addlegendimage{green!50!black, mark=diamond*}\addlegendentry{wall line, boundary values, max}
\addplot[gray, domain=16:512, samples=2, forget plot] {8/x}; \addplot[gray, domain=16:512, samples=2, forget plot] {1.6/x^2};
\node[font=\scriptsize, gray, anchor=south] at (axis cs:24,0.36) {$N^{-1}$}; \node[font=\scriptsize, gray, anchor=north east] at (axis cs:200,0.00004) {$N^{-2}$};
\nextgroupplot[title={3D, oblique slice}]
\addplot[red, mark=*] table[x=N, y=nodeint_l2] {data/conv3d.dat};
\addplot[red, dashed, mark=o] table[x=N, y=nodeint_linf] {data/conv3d.dat};
\addplot[blue, mark=square*] table[x=N, y=node_bnd_linf] {data/conv3d.dat};
\addplot[black!70, mark=triangle*] table[x=N, y=wall_cell] {data/conv3d.dat};
\addplot[green!50!black, mark=diamond*] table[x=N, y=wall_fixed] {data/conv3d.dat};
\addplot[gray, domain=8:64, samples=2] {4/x}; \addplot[gray, domain=8:64, samples=2] {1.2/x^2};
\node[font=\scriptsize, gray, anchor=south] at (axis cs:12,0.36) {$N^{-1}$}; \node[font=\scriptsize, gray, anchor=north east] at (axis cs:30,0.0011) {$N^{-2}$};
\end{groupplot}
\node[anchor=north] at ($(group c1r1.south)!0.5!(group c2r1.south) + (0,-1.0cm)$) {\pgfplotslegendfromname{convleg}};
\end{tikzpicture}
\caption{Interpolation errors against the exact manufactured fields on smoothly distorted hexahedral meshes with $N$ cells per direction ($h=1/N$). Interior: mesh points (2D) or slice points (3D) more than $1.5h$ from the boundary. Grey lines: slopes of first and second order. In 2D the boundary values on the wall are exact to the written digits ($\numWallFixedTwoMax$, not shown).}
\label{fig:conv}
\end{figure}

The results are summarised in \cref{fig:conv}. For the tested smoothly distorted meshes, the interior point values converge at approximately second order: the observed RMS order over the finest meshes is \numOrdTwoDint\ in 2D and \numOrdThreeDint\ on the oblique 3D slice (maximum norm \numOrdThreeDintInf\ in 3D). Boundary points, in contrast, converge at first order (\numOrdTwoDbnd\ and \numOrdThreeDbnd), as expected for the one-sided average \eqref{eq:c2p}. Because the boundary points constitute a fraction $O(h)$ of all points, the RMS over all points accordingly behaves as $h^{3/2}$ (observed \numOrdTwoDall\ and \numOrdThreeDall).

Along the wall, the wall-cell values are first-order approximations of the wall values (observed \numOrdWallCellTwo\ and \numOrdWallCellThree), whereas the boundary values given in the field file are returned exactly: in 2D, the largest difference is $\numWallFixedTwoMax$, which corresponds to the ten significant digits of the line export. In 3D, the line along $z=\frac12$ crosses each wall face away from its centre. The face value, which belongs to the face centre, therefore differs from the exact field at the crossing point by $O(h)$ (observed order \numOrdWallFixedThree), although the returned values are still exactly the solver's face values.

\subsection{Accuracy of exported data}
\label{sec:roundtrip}
Data exported from \HP\ should contain the same values as the solver's files. This was checked in two ways. First, the cell values of the 2D test cases were exported to CSV files with 17 significant digits. This is the precision needed to write a double-precision number in decimal form without loss: a value written in this way and read again is exactly the original number. Compared with the exact test field evaluated at the exported cell centres, the exported values differ by at most $\numCellExportMax$, which is the size of the rounding error in the computed cell centres. Second, VTK and zoned ASCII files written by \HP, with the values stored either at the cells or at the mesh points, were opened again in \HP. They were displayed on the same polygons as the original data, and the plots were visually identical.

\subsection{Performance}
\Cref{tab:perf} reports the time required to read a case (mesh and fields) and the time required to redraw the plot after a change of colormap, for which the interpolation is cached, and after the first change to another field, which requires a new interpolation. A redraw is timed from the change of the setting until the updated frame has been drawn. Every measurement was repeated with five fresh page loads, so that no interpolation was cached from an earlier run, and the medians are reported. Loading takes about \numLoadPerFace~$\mu$s per mesh face on the larger meshes and grows linearly with mesh size, so that the largest test case, with \numBigCells\ cells and one million faces, loads in \numBigLoad~s. A change of colormap redraws in at most \numRestyleMax~ms, and a change of field, including the new interpolation, in at most \numNewFieldMax~ms; the tool therefore stays interactive at these sizes. The production cases of \cref{tab:reader} load in \numProdLoadMin--\numProdLoadMax~s. In practice, memory rather than speed is the limiting factor, since a browser tab can use at most a few gigabytes; cases larger than those reported here have not been tested systematically.

\begin{table}[!htbp]
\centering
\caption{Timings on the synthetic meshes (Chrome~156, AMD Ryzen~7 5700G). Medians of five page loads. Restyle: change of colormap (cached interpolation); new field: first change to another field (new interpolation); in 3D the main view shows the 3D scene with the oblique slice of \cref{sec:mms}; ``walls'': the same redraw with all six domain walls colored.}
\label{tab:perf}
\small
\begin{tabular}{@{}l r r r r r r r@{}}
\toprule
 & Cells & Faces & Load (ms) & $\mu$s/face & Restyle (ms) & New field (ms) & Walls (ms) \\
\midrule
\csname @@input\endcsname data/perf_table.tex
\bottomrule
\end{tabular}
\end{table}

\section{Applications}
\label{sec:applications}

The following examples illustrate how the operations of \cref{sec:algorithms} combine in practice. Each figure was produced entirely within \HP\ and exported directly from it, without subsequent editing.

\subsection{Impinging supersonic jet}
\label{sec:jet}
\begin{figure}[!htbp]
\centering
\includegraphics[width=0.55\textwidth]{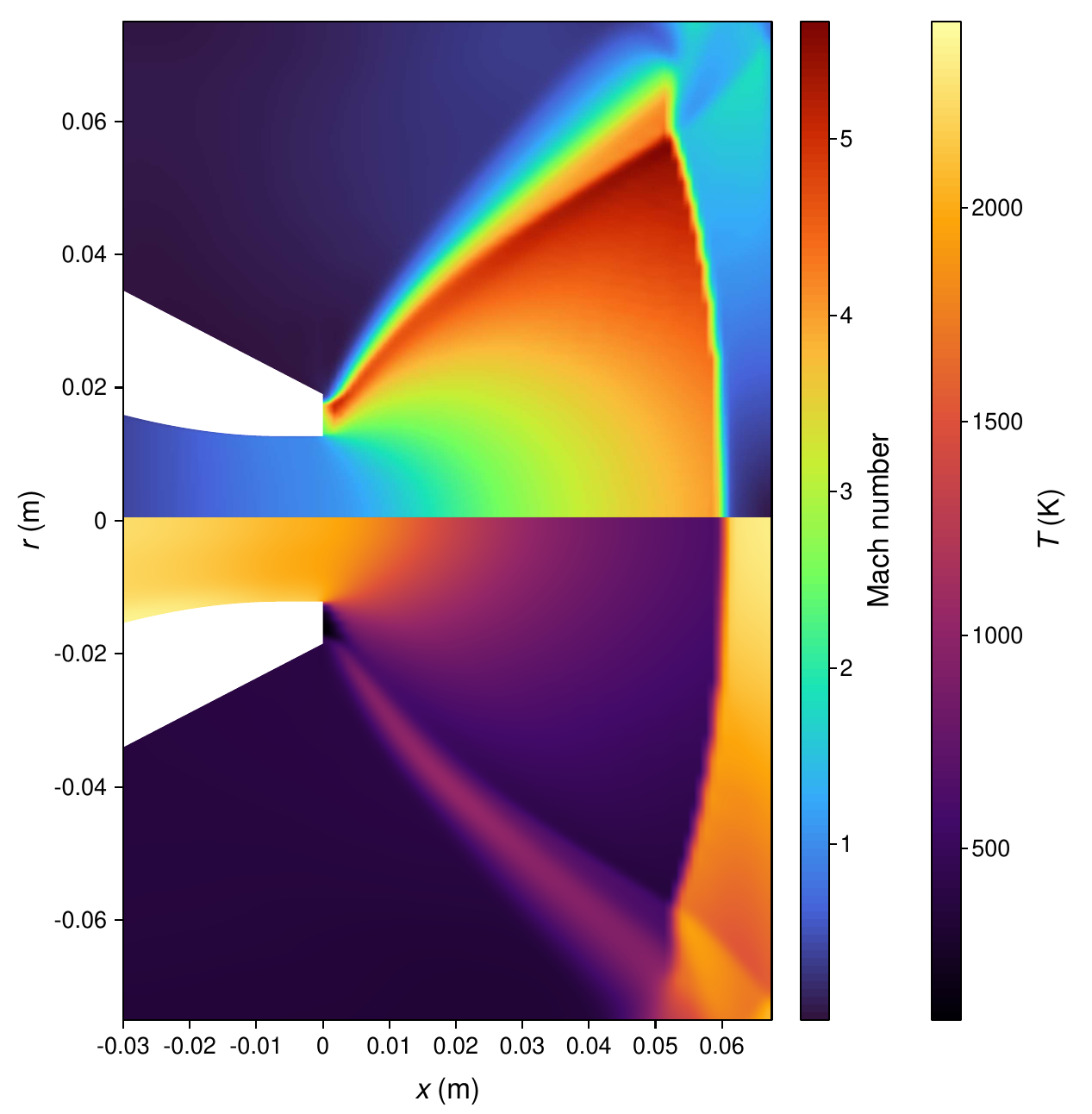}
\caption{Impinging supersonic jet (axisymmetric, 68\,123 cells). Upper half: Mach number computed in the calculator; lower half: temperature, mirrored about the axis with its own colormap. Smooth shading; vector PDF exported by \HP.}
\label{fig:jet}
\end{figure}

\Cref{fig:jet} shows an underexpanded axisymmetric jet with a nozzle pressure ratio of 4 impinging on a flat plate. At this pressure ratio the flow leaves the nozzle at a pressure above ambient and expands further outside it; the expansion waves reflect from the jet boundary as compression waves that coalesce into a barrel shock, and a second, nearly normal shock forms ahead of the plate, behind which the gas is decelerated into a hot stagnation region. The case was opened directly from its folder and recognized as axisymmetric from its two \texttt{wedge} patches. The Mach number was defined in the calculator as \texttt{U\_mag/sqrt(1.4*287*T)} and is shown above the axis, while the temperature is displayed in the mirrored lower half with its own colormap and colorbar. In this way the expansion of the jet, its boundary shock, the stand-off shock ahead of the plate, the cold core of the jet and the hot stagnation region on the plate are all visible in a single figure, which was exported as a vector PDF.

\subsection{Hypersonic flow over a cylinder}
\begin{figure}[!htbp]
\centering
\begin{subfigure}[b]{0.36\textwidth}\centering\includegraphics[width=\textwidth]{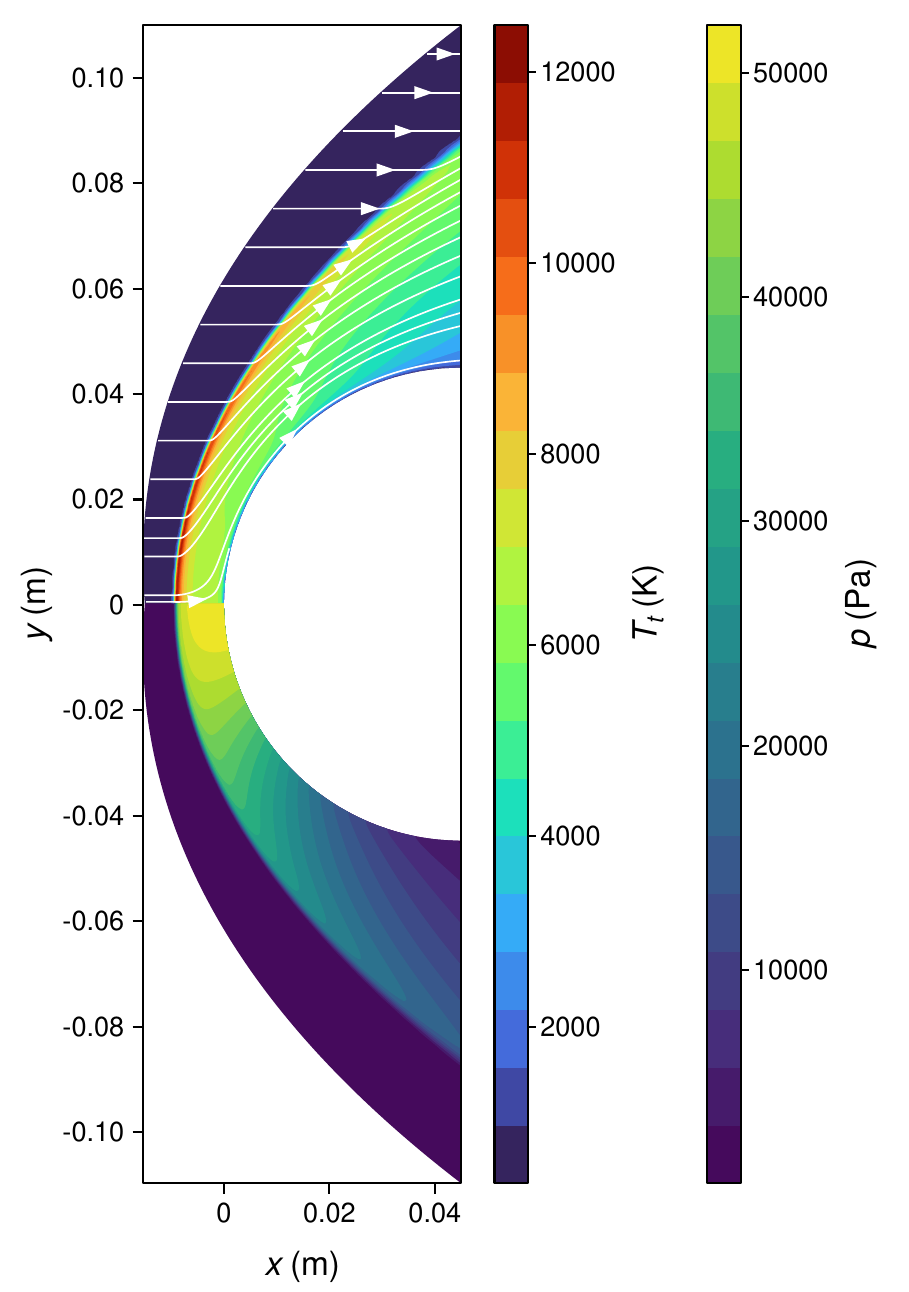}\caption{}\end{subfigure}\hfill
\begin{subfigure}[b]{0.62\textwidth}\centering\includegraphics[width=\textwidth]{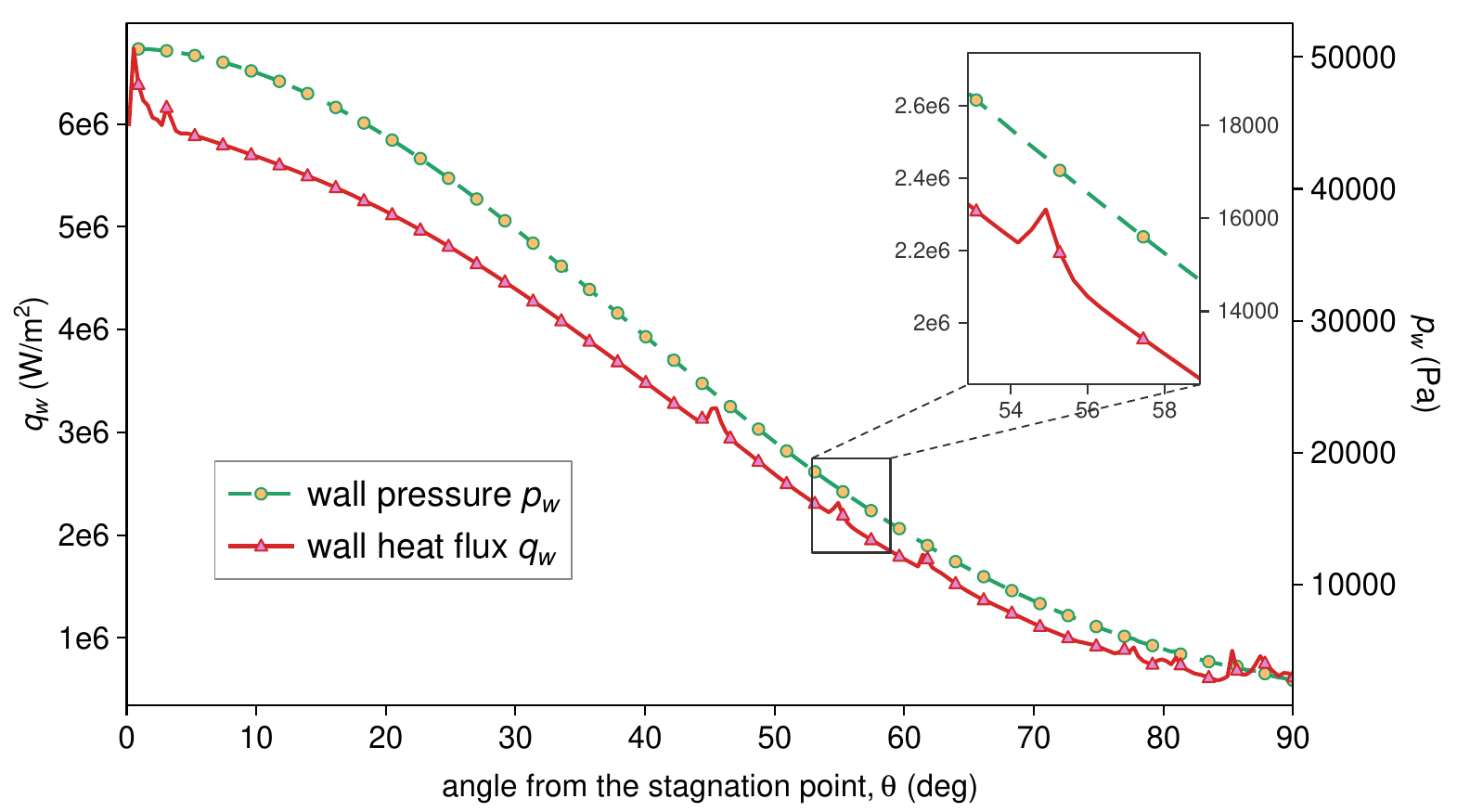}\caption{}\end{subfigure}
\caption{Hypersonic two-temperature flow over a cylinder (62\,001 cells). (a)~Translational--rotational temperature with streamlines (upper half) and pressure (mirrored lower half). (b)~Wall heat flux and wall pressure (right axis) along the cylinder, extracted from the boundary patch.}
\label{fig:heg}
\end{figure}

In hypersonic flow over a blunt body, a strong bow shock stands ahead of the body, and the gas in the thin shock layer is heated to temperatures at which vibrational excitation, dissociation and, eventually, ionization become important. Since these processes are slower than translational relaxation, the shock layer is in thermochemical nonequilibrium, and the solution is described by a translational--rotational and a vibrational--electronic temperature together with the mass fractions of eleven species of air. Owing to the symmetry of the problem, the flow was computed on half of the domain. In \cref{fig:heg}(a) the translational--rotational temperature is shown with automatically seeded streamlines in the upper half, and the pressure is mirrored into the lower half; the stand-off distance of the bow shock and the deflection of the streamlines across it are clearly resolved. \Cref{fig:heg}(b) shows the wall heat flux and the wall pressure along the cylinder, which were extracted from the wall patch using the boundary values written by the solver (\cref{sec:surface}). The angle from the stagnation point was computed in the calculator from the extracted coordinates as \texttt{deg(atan2(y, max(x) - x))}, and the pressure is plotted on a second axis. Both quantities peak at the stagnation point and decrease monotonically along the surface, as expected for a cylinder in hypersonic flow. The small steps in the heat flux are present in the solver's wall data; the extraction returns the face values unchanged and, deliberately, does not smooth them.

\subsection{Hybrid particle--continuum cold-spray nozzle}
\label{sec:hybrid}
\begin{figure}[!htbp]
\centering
\includegraphics[width=0.92\textwidth]{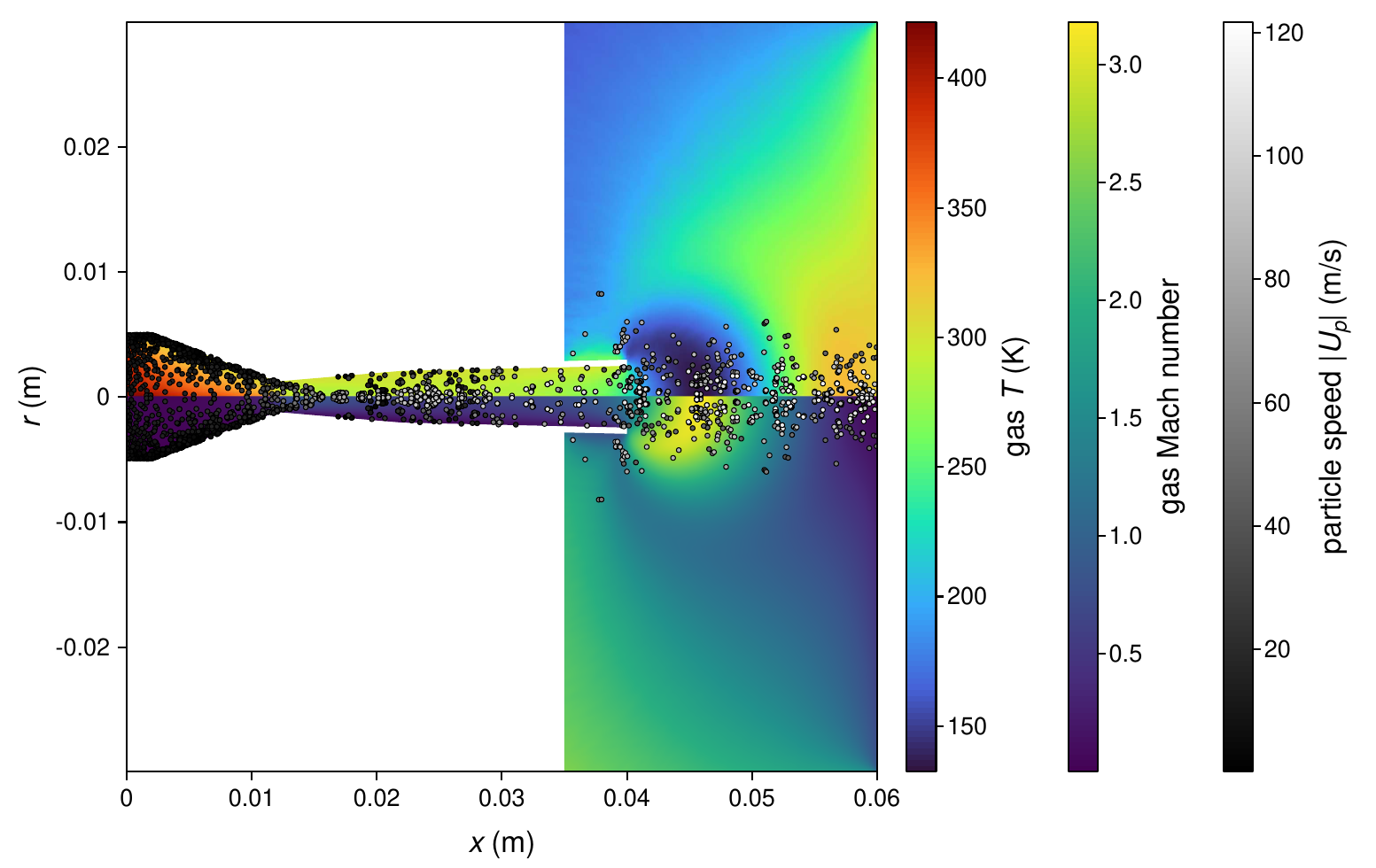}
\caption{Axisymmetric cold-spray nozzle computed with a hybrid DSMC/MP-PIC solver (24\,980 cells). Gas temperature (upper half) and gas Mach number (mirrored lower half), with the solid particles colored by speed.}
\label{fig:hybrid}
\end{figure}

In cold spraying, solid powder particles are accelerated by a supersonic gas flow through a converging--diverging nozzle and deposited on a substrate. Downstream of the nozzle exit the gas expands into a low-pressure chamber where it becomes rarefied, so that a continuum description is no longer adequate; the gas was therefore simulated with DSMC, and the solid particles with MP-PIC, both of which store their particles as Lagrangian clouds. \HP\ lists the two clouds of the case together with their sizes (1\,887 solid particles and 1\,178\,684 gas simulator particles at the time shown) and, by default, reads only the smaller cloud. In \cref{fig:hybrid} the solid particles are drawn over the gas temperature and colored by their speed, while the gas Mach number is mirrored into the lower half. The figure thus relates the particle velocity directly to the gas flow that accelerates it: the particles gain speed in the diverging section, where the gas expands and cools, and retain their momentum in the chamber, where the gas is much slower to respond. Hovering over a particle displays all of its properties (temperature, velocity, diameter, phase state and so on); these values were checked against the cloud files. The particles can, moreover, be listed as a table and exported.

\subsection{Three-dimensional fields: slices, walls and isosurfaces}
\label{sec:3dapps}
\begin{figure}[!htbp]
\centering
\includegraphics[width=0.78\textwidth]{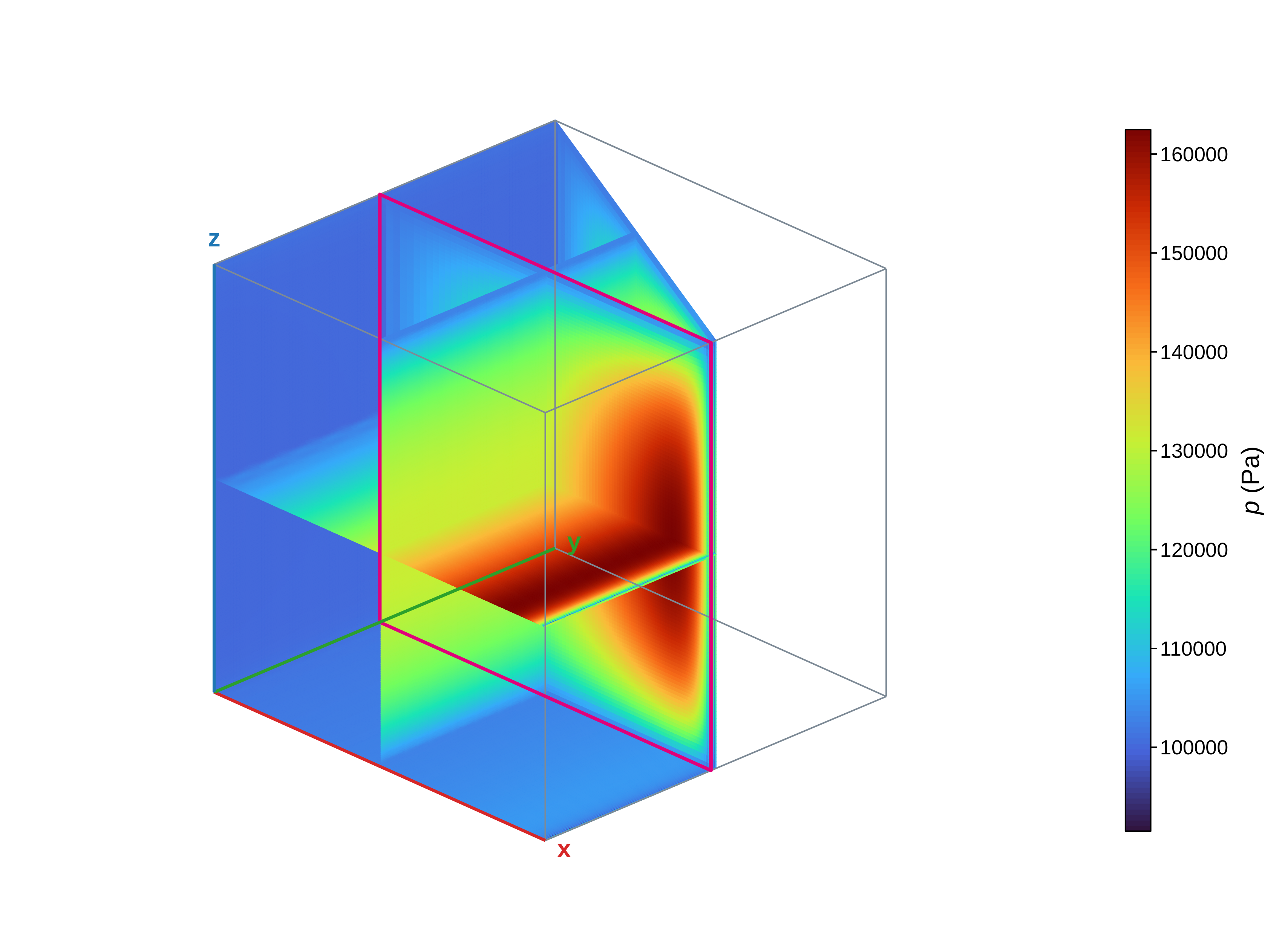}
\caption{Pressure in the three-dimensional gas film of a bearing on three slice planes and the back walls of the domain (157\,080 cells; axes scaled independently).}
\label{fig:bearing}
\end{figure}

\Cref{fig:bearing} shows the pressure in the three-dimensional gas film of a self-acting slider bearing (157\,080 cells) on three orthogonal slice planes together with the back walls of the domain. The film is $10~\mu$m long and $10~\mu$m wide, and its height decreases from $199$~nm at the inlet to half this value at the outlet; the lower surface moves at $33$~m/s, so that the gas is dragged into the converging gap and compressed (compressibility number $\Lambda=5$). Because the film is about fifty times thinner than it is long, the axes are scaled independently; otherwise the pressure variation across the film thickness could not be seen. This variation is small over most of the film (median $0.1\%$ of the overall pressure difference) but reaches about $11\%$ near the outlet, where the gap is narrowest.

\begin{figure}[!htbp]
\centering
\includegraphics[width=0.62\textwidth]{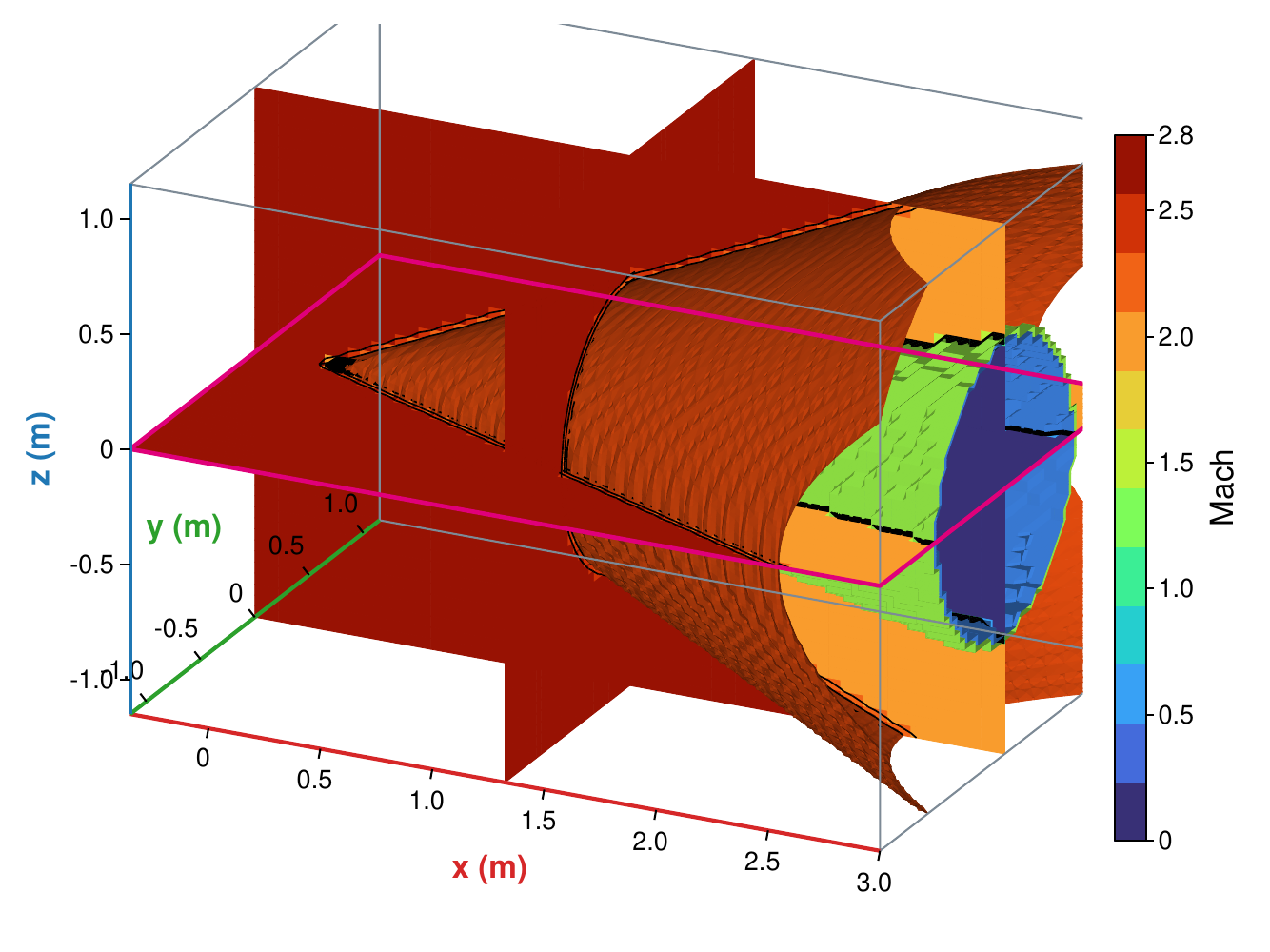}
\caption{Supersonic flow past a cone: Mach number on a $65 \times 49 \times 49$ structured grid, shown on the slice $z = 0$ (plane 25 of 49) together with isosurfaces at $M = 0.467$, $1.4$ and $2.333$ (89\,132 triangles in total).}
\label{fig:cone}
\end{figure}

The second three-dimensional example, \cref{fig:cone}, concerns supersonic flow past a cone on a structured $65\times49\times49$ grid and illustrates the combination of slice planes with isosurfaces. The Mach number is shown on the plane $z=0$, which is the 25th of the 49 grid planes and therefore contains the axis of the cone, together with three isosurfaces at $M=0.467$, $1.4$ and $2.333$. These values divide the Mach-number range $[0,2.8]$ into three equal intervals and lie at their centres, so that each isosurface represents one part of the flow. The highest level follows the conical shock attached to the apex of the cone; because the flow behind an attached conical shock is nearly uniform along rays from the apex, this isosurface is itself very nearly a cone. The two lower levels enclose, respectively, the region in which the flow has been decelerated to low supersonic or subsonic speed and the slow core of the wake behind the base of the cone. Together, the three isosurfaces consist of 89\,132 triangles extracted by marching tetrahedra (\cref{sec:algorithms}); they close without holes, and the staircase pattern visible on the lower levels reflects the resolution of the sampling lattice rather than an error of the extraction. The intersection of the isosurfaces with the slice plane, where colors of the plane and of the surfaces meet, confirms that both are computed from the same interpolated field. Even with this number of primitives, the scene is depth-sorted and redrawn interactively by the painter's algorithm and can be exported as a vector file.

\subsection{Color scales for fields with a large dynamic range}
\begin{figure}[!htbp]
\centering
\begin{subfigure}[b]{0.79\textwidth}\includegraphics[width=\textwidth]{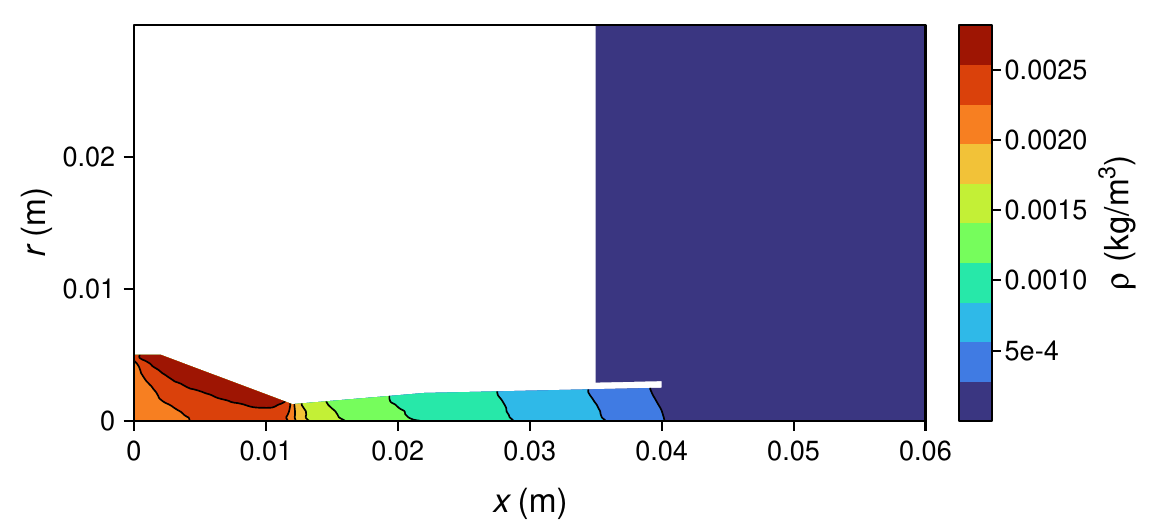}\caption{linear, 10 levels}\end{subfigure}\hfill
\begin{subfigure}[b]{0.79\textwidth}\includegraphics[width=\textwidth]{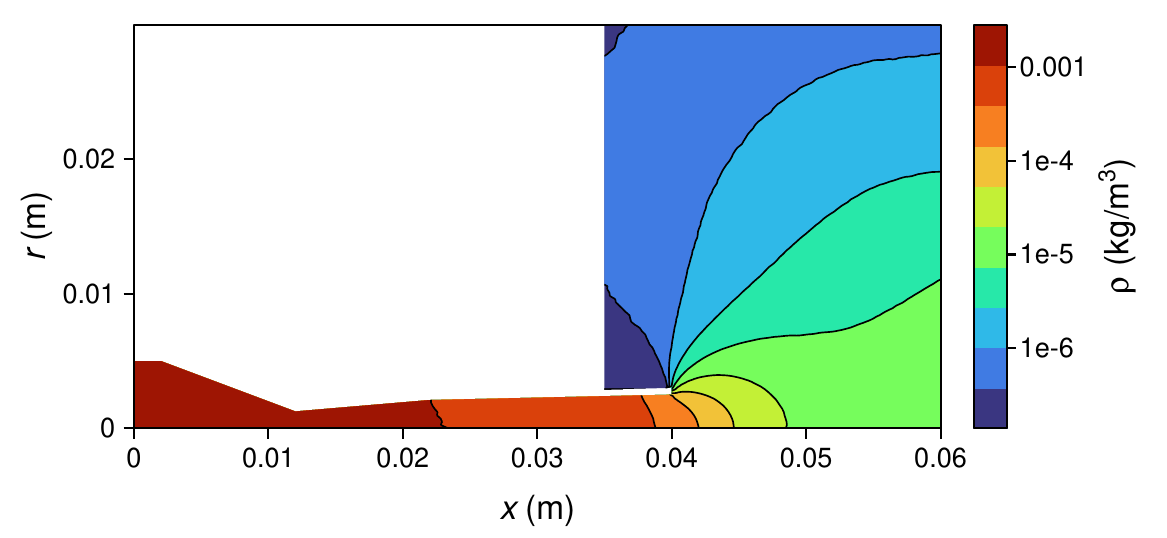}\caption{exponential, 10 levels}\end{subfigure}\\[1ex]
\begin{subfigure}[b]{0.79\textwidth}\includegraphics[width=\textwidth]{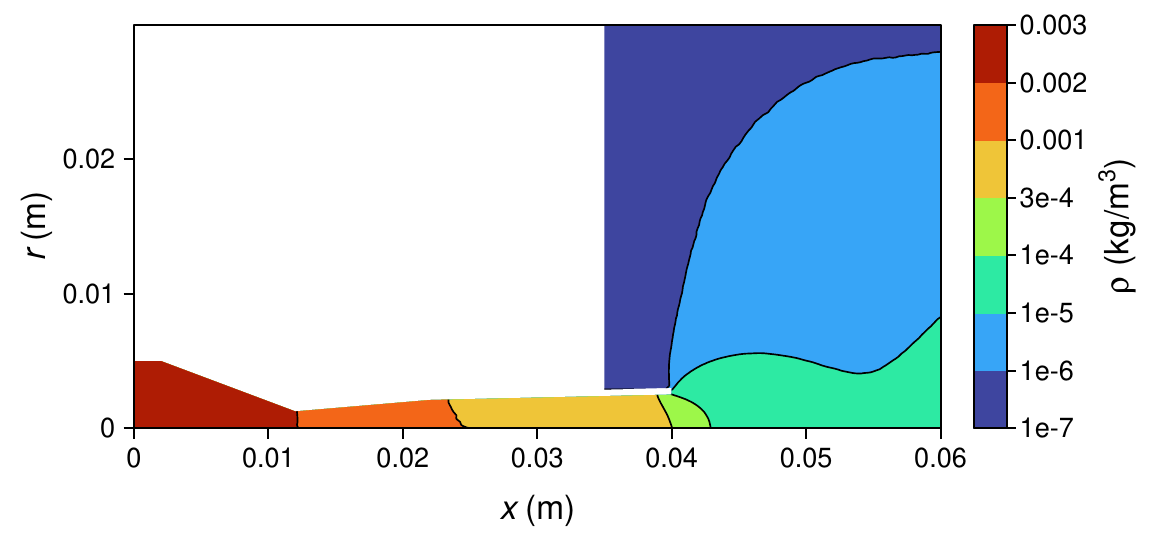}\caption{manual: $10^{-7}$, $10^{-6}$, $10^{-5}$, $10^{-4}$, $3{\times}10^{-4}$, $10^{-3}$, $2{\times}10^{-3}$, $3{\times}10^{-3}$}\end{subfigure}
\caption{Gas density of the cold-spray case with linear, exponential and manual level spacing (\cref{eq:levels}). Bands, contour lines and colorbar follow the same scale.}
\label{fig:levels}
\end{figure}

Many fields of practical interest span several orders of magnitude, and their appearance then depends strongly on the color scale. \Cref{fig:levels} compares the three level spacings of \cref{sec:levels} on the gas density of the cold-spray case, which varies by four orders of magnitude between the nozzle and the expansion chamber. With linear levels the entire plume falls into the lowest band and its structure is lost. Exponential levels, by contrast, resolve the plume, because each band now corresponds to a fixed ratio of densities. Manual levels, finally, allow the user to place the band boundaries at physically meaningful values, here with finer resolution in the range in which the flow becomes rarefied.

\section{Limitations}
\label{sec:limitations}
\HP\ is designed for interactive work on a single computer. The case must fit into the memory of a browser tab, and decomposed cases must be reconstructed first.  The 3D view relies on the painter's algorithm on the CPU rather than on GPU rendering; it can therefore misorder primitives that intersect or overlap cyclically, and it slows down for scenes with many hundreds of thousands of polygons. Interpolated point values are first-order accurate at boundary points (\cref{sec:mms}), as in other tools that use cell-to-point averaging, and wall lines report face values, which belong to the face centers. Particles of clouds stored only in barycentric coordinates are placed at their cell centers. Although the VTK reader accepts the output of finite-element codes, element-specific features such as higher-order shape functions, integration-point data or shell and beam elements are not interpreted; such data are currently treated as linear cells with point or cell values. Finally, \HP\ is free to use, but its source code is not open. 

\section{Conclusions}
This study introduced and quantitatively verified HAVAPlot, a scientific visualization framework that runs entirely on the user's computer and enables installation-free post-processing of numerical simulation data directly within a web browser. The framework integrates native simulation-data parsing, mesh-based interpolation, multidimensional visualization, and publication-quality graphical output without requiring external data transfers. The work demonstrates that the everyday post-processing of simulation results, from opening a case as written by the solver to publication figures and exported data, can be carried out without installation and without uploading data, and that this convenience need not come at the expense of correctness.

Two conclusions can be drawn from this work. First, every value that \HP\ displays or exports is either the solver's own value or a well-defined interpolation of it, and it is shown at the position where the solver defines it. By reproducing the solver's own geometric decomposition, \HP\ places every cell value at the position the solver assigns to it. The tests confirmed this: the cell geometry agreed with an independent implementation of OpenFOAM's algorithm to rounding error, \numTotalValues~million field values were reproduced to machine precision, and the interpolation was second-order accurate in the interior of 2D meshes and on 3D slices. Second, because all processing takes place on the user's own computer and the data are never uploaded to an external server, the tool can facilitate work with unpublished or proprietary results, subject to applicable institutional and data-security requirements. This is particularly useful in teaching, collaborative work, and environments where software installation is restricted.

The applications show that a single tool covers a wide range of problems, from axisymmetric compressible flows with mirrored multi-field displays, through wall data extracted from boundary patches and Lagrangian particle clouds of hybrid particle--continuum simulations, to three-dimensional fields with slices and isosurfaces. Although all examples in this paper originate from fluid dynamics, nothing in the pipeline is specific to it: the data-set abstraction, the slicing and interpolation algorithms and the color scales apply equally to the displacement, strain and stress fields of structural analyses or to the temperature fields of heat-transfer problems, which enter through the VTK reader.

Future work will address the limitations identified above. Planned developments include GPU rendering of large 3D scenes, which would remove both the ordering errors of the painter's algorithm and its cost for very large scenes; the direct reading of decomposed and dynamic-mesh cases; boundary-aware interpolation that restores second-order accuracy at boundary points by using the boundary values where the solver provides them; and native support for finite-element data, including higher-order elements and integration-point quantities, together with verification cases from structural mechanics.

\section*{Code and data availability}
\HP\ is available free of charge at \url{https://www.havalab.org/handy-tools/} (Tools $\rightarrow$ Data Visualizer) and uses only standard web interfaces; the tests reported here were run in Google Chrome. The ancillary files of this article contain the generator of the synthetic test cases, the independent Python reader, the browser test scripts and the raw results.
\section*{Use of AI tools}
Generative AI (Claude, Anthropic) was used to assist with parts of the software development and the test scripts.

\section*{Acknowledgements}
The author gratefully acknowledges the financial support of Rensselaer Polytechnic Institute.

\bibliographystyle{unsrtnat}
\bibliography{refs}

\end{document}